\documentclass[preprint,aps]{revtex4}%
\usepackage{amsfonts}
\usepackage{amsmath}
\usepackage{amssymb}
\usepackage{graphicx}%
\begin{document}
\preprint{ }
\title[Klein Paradox and Graphene]{1-D Dirac equation, Klein Paradox and Graphene}
\author{S. P. Bowen}
\affiliation{Chicago State University, Chicago, IL 60628}

\begin{abstract}
Solutions of the one dimensional Dirac equation with piece-wise constant
potentials are presented following standard methods. \ These solutions show
that the Klein Paradox is non-existent and represents a failure to correctly
match solutions across a step potential. \ Consequences of this exact solution
are studied for the step potential and a square barrier. \ Characteristics of
massless Dirac states and the momentum linear band energies for Graphene are
shown to have quite different current and momentum properties.

\end{abstract}
\volumeyear{year}
\volumenumber{number}
\issuenumber{number}
\eid{identifier}
\date[Date text]{date}
\received[Received text]{date}

\revised[Revised text]{date}

\accepted[Accepted text]{date}

\published[Published text]{date}

\startpage{1}
\email{sbowen@csu.edu}
\maketitle

\section{Introduction}

The renewed interest in graphene\cite{graphene} and the close analogy of its
band structure to the spectrum of the zero mass Dirac equation suggests that a
re-examination of several aspects of the one dimensional Dirac equation should
be carried out. \ The first of these aspects is the well known Klein
paradox\cite{KleinParadox} which continues to persist in the literature. \ The
second of these aspects is the question of how closely the graphene spectrum
resembles the Dirac spectral properties and states. \ Dragoman in an excellent
paper\cite{Budapest} has recently examined some of these issues and has noted
that there is no Klein Paradox. \ In this paper we will examine the
mathematics of solutions to the one dimensional Dirac equation in the presence
of a piece-wise constant potential step (the Klein problem). \ The solutions
of this differential equation are exact and can be carried out analytically.
\ The usual method of analysis will involve the delta function normalization
of wave functions in the continous spectrum. \ In order to clarify some of the
mathematical details a slightly more meticulous method of calculation
originally suggested by Von Neumann\cite{VonNeumann} will be followed so that
all of the mathematical inferences can be carried out transparently. \ In the
first section details of the solutions of the differential equation will be
discussed. \ At the end of that section the wave function solutions for the
potential step will be exhibited. \ This will show that the Klein paradox
arises because of a mis-application of the solutions of piece-wise constant
potentials for these one dimensiional Dirac equations. \ In the next section
will be a short discussion of the admixture of free negative energy states
into the positive energy states for this step potential system if the
potential were to be turned on instantaneously. \ This admixture should
monitor the creation of electron-positron pairs near the potential step.
\ After this admixture discussion a characteristic of orthogonality in the
thermodynamic limit will be discussed. \ The reflection coefficient for finite
width square potential barriers will be displayed and finally the effect of
piece-wise constant potential steps on zero mass states will be discussed as
well as a comparison with both the energy spectrum and current densities for
\ a "one dimensional" Graphene.

\section{Solutions of Piece-wise Constant Potential Differential Equations}

One dimensional solutions of the Schrodinger and Dirac equations essentially
reduce to finding the finite, continous, and differentiable solutions of a
Sturm-Liouville type of differential equation on the whole line interval
$(-\infty,\infty)$\cite{MessiahandMerzbacher},\cite{Messiah}$.$ \ For the
discrete spectrum of such equations there is an orthonormal set of
eigenstates. \ For the continuous spectrum the momemtum eigenstates are
functions that can only be normalized with Dirac delta functions. \ The delta
function normalization is generally quite satisfactory except for some
calculations involving squares of delta functions. \ 

Because of a desire to be mathematically careful and transparent in this
examination of the Klein paradox, a procedure will here be followed that dates
back to von Neumann\cite{VonNeumann} and many others\cite{Others}.
\ Specifically, all of the wave functions to be used here will be "box
normalized" in regions of length $L$ and final results will be obtained in the
thermodynamic limit ($L\rightarrow\infty)$. \ Within the interval of length
$L$ the allowed momentum values will be discrete $k=2\pi n/L,$ where $n$ is an
integer. \ This whole process will not be essential for the demonstration that
the Klein paradox is not a paradox at all, but will be useful in studying the
admixture of free particle states and those interacting states in the presence
of the potential step. \ For each finite $L$ we will find that the eigenvalue
spectrum will be discrete and that in the thermodynamic limit $(L\rightarrow
\infty)$ the spectrum becomes continuous and the eigenstates become delta
function normalized. \ Details of the transition to the thermodynamic limit
have been discussed in detail by Messiah\cite{Messiah}, Arfken and
Weber\cite{Arfken}, and Sneddon\cite{Sneddon} and some of these details will
be discussed in the following.

The Dirac equation with a constant potential has exact solutions which are the
same as the free particle solutions except that the energy $E_{k}$ can be
different from the free particle case%
\begin{equation}
E_{k}=V_{0}\pm\sqrt{m^{2}+k^{2}}%
\end{equation}
by the addition of the constant potential. \ In order to simplify the notation
in this study the velocity of light $c=\hbar=1$ will be used in the formulas
to follow. \ The Dirac equation will be studied with 2x1 spinors for
simplicity. \ This is a standard simplification sometimes described as "no
spin" or "single spin". \ The basic idea of this paper is that we consider
these solutions to Dirac equations in two regions $(-L,0)$ and $(0,L)$ and
construct all wave functions out of two parts%
\begin{equation}
\Psi(z)=\psi_{L}(z)\Theta(-z)+\psi_{R}(z)\Theta(z)
\end{equation}
where $\Theta(z)$ is the Heaviside function (unit step function at zero),
$\psi_{L}(z)$ is a solution in the left side interval $(-L,0)$ and $\psi
_{R}(z)$ is a solution in the right interval $(0,L).$ \ The boundary condition
is that the two solutions be continuous at the origin%
\begin{equation}
\psi_{L}(0^{-})=\psi_{R}(0^{+}).
\end{equation}

In each region, $(-L,0)$ and $(0,L),$the orthonormal positive and negative
energy states, respectively, are given by%
\begin{equation}
p_{k}(z)=\frac{1}{\sqrt{L}}\frac{1}{\sqrt{1+u_{k}^{2}}}\left(
\begin{array}
[c]{c}%
1\\
u_{k}%
\end{array}
\right)  e^{ikz}%
\end{equation}
and%
\begin{equation}
n_{-k}(z)=\frac{1}{\sqrt{L}}\frac{1}{\sqrt{1+u_{k}^{2}}}\left(
\begin{array}
[c]{c}%
u_{k}\\
1
\end{array}
\right)  e^{-ikz}%
\end{equation}
where%
\begin{equation}
u_{k}=\frac{k}{m+\sqrt{m^{2}+k^{2}}}.
\end{equation}

Both of the orthonormal solutions above are written for a positive-direction
probability current density $J_{k}$ which is given by%
\begin{equation}
J_{k}=\frac{2u_{k}}{1+u_{k}^{2}}.
\end{equation}
\ Reversing the $k$ value gives the current in the negative direction carrying
states. \ The energies of these states will depend on the side of the origin.
\ On the left the energy will be $\pm\sqrt{m^{2}+k^{2}}$ and on the right
$V_{0}\pm\sqrt{m^{2}+k^{\prime2}}.$ \ In the following paragraphs the wave
vectors on the right side $(z\geq0)$ will be indicated with a prime as
$k^{\prime}.$ \ 

On either side of the origin these solutions are orthonormal%
\begin{equation}
\langle n_{k}|n_{k^{\prime}}\rangle=\langle p_{k}|p_{k^{\prime}}\rangle
=\Delta_{k,k^{\prime}}=\frac{e^{i(k-k^{\prime})L}-1}{i(k-k^{\prime})L}%
\end{equation}
where $\Delta_{k,k^{\prime}}$ is the Kronecker delta, $\Delta_{k,k^{\prime}%
}=1$ if $k-k^{\prime}$ and zero if $(k-k^{\prime})L=2\pi(n-n^{\prime})\neq0.$
\ In each interval the allowed momentum values are $k=2\pi n/L,$ where $n$ is
any integer. \ Similarly, $\langle n_{k}|p_{k^{\prime}}\rangle=0.$ \ The
discrete $k$ values are those which guarantee that the momentum operator is
hermitian, namely, that $\psi^{\dagger}(0)\psi(0)=\psi^{\dagger}(L)\psi(L)$
for the right side, and similarly on the left side. \ The usual delta function
nromalization is achieved by removing the normalization factor $1/\sqrt{L}$
from each wave function and recalling that for all sequences leading to the
thermodynamic limit%
\begin{equation}
\lim_{L\rightarrow\infty}L\Delta_{k,k^{\prime}}=\delta(k-k^{\prime})
\end{equation}
where $\delta(k-k^{\prime})$ is the Dirac delta function$.$ \ In the following
it will often be the case that the normalizations for each half interval
$1/(\sqrt{L}\sqrt{1+u_{k}^{2}})$ will be discarded and normalization on the
whole interval $(-L,L)$ will be re-calculated.

For the Dirac equation on the whole interval $(-\infty,\infty)$ there are no
states in the mass gap $-m\leq E\leq m.$ \ But, for each of the two half
intervals there are evanescent (exponential) wave functions that preserve
momentum hermiticity. \ These are labeled by a complex momentum $i\kappa$ and
have both positive and negative energy relative to the center of the mass gap
$V_{0}\pm\sqrt{m^{2}-\kappa^{2}}$ and can be written \ for each fixed $L$ in
the sequence as%

\begin{equation}
O_{\kappa}^{(>+)}(z)=\left(
\begin{array}
[c]{c}%
1\\
iw_{\kappa}%
\end{array}
\right)  e^{-\kappa z}+e^{-\kappa L}\left(
\begin{array}
[c]{c}%
1\\
-iw_{\kappa}%
\end{array}
\right)  e^{\kappa z}%
\end{equation}%
\begin{equation}
O_{\kappa}^{(>-)}(z)=\left(
\begin{array}
[c]{c}%
-iw_{\kappa}\\
1
\end{array}
\right)  e^{-\kappa z}+e^{-\kappa L}\left(
\begin{array}
[c]{c}%
iw_{\kappa}\\
1
\end{array}
\right)  e^{\kappa z}%
\end{equation}
where
\begin{equation}
w_{\kappa}=\frac{\kappa}{m+\sqrt{m^{2}-\kappa^{2}}},
\end{equation}
for the interval $(0,L).$ Similar functions $O_{\kappa}^{(<+)}$ and
$O_{\kappa}^{(<-)}$ are defined for the interval $(-L,0).$ \ In the
thermodynamic limit the second term vanishes and these functions yield the
usual single exponential function.

When the step potential $V_{0}$ is zero, there are no allowed states in the
mass gap. Even when $0\leq V_{0}\leq2m$ there are no states in the part of the
gap which remains on both sides of the origin.

The general solution is now achieved by equating at the origin the two
functions, one from from each side at the origin, that have the same energy,
and using this equality of energy to determine the relationship between the
momenta on the two sides. \ In Tables (1 and 2) are shown the energy ranges
and the structure of the wave functions in each region which have the same
energy. \ Note that for simplicity, we are examining states that initiate with
a current from the left and contain a reflected wave on the left and
transmitted wave on the right.

There are two distinct $V_{0}$ energy ranges and to simplify the discussion
Table 1 shows the wave function pairings and the relationships between the
left and right side momenta for small $0\leq V_{0}\leq2m$ and Table 2 shows
the same for $V_{0}>2m.$ \ The table shows the wave functions and the
relationship between the momenta on each side of the origin and for each
energy range.

Referring to Table I, for small $V_{0}$ $(0\leq V_{0}\leq2m),$ we will discuss
the states from the top of the Table (highest energy) to the bottom (lowest
energy). \ At the top are the states for which $E_{k}>V_{0}+m.$ \ As can be
seen in the Table we have "positive energy" functions $p_{k}(z)+f_{k}$
$p_{-k}(z)$ on the left ($f_{k}$ is the reflection amplitude) and on the right
we have $g_{k}p_{k^{\prime}}(z)$ where $g_{k}$ is the transmission amplitude.
\ The next set of functions are in the energy interval $m\leq E_{k}\leq
V_{0}+m$, which are of the form $p_{k}(z)+e^{i\phi}p_{-k}(z)$ in the left side
and are matched with an evanescent wave $O_{\kappa}(z)$ on the right side.
\ Because the evanescent states carry no current, the reflection amplitude has
unit magnitude and thus gives complete reflection. \ The next energy range
$V_{0}-m\leq E\leq m$ corresponds to the empty mass gap with no wave functions
satisfying the boundary condition. \ The next lower energy range $-m\leq
E_{k^{\prime}}\leq V_{0}-m$ corresponds to another combination of which half
is evanescent and cannot carry any current. \ These states have an evanescent
wave $O_{-\kappa}(z)$on the left and $n_{k}(z)+e^{i\phi}n_{-k}(z)$ on the
right. \ The reflection amplitude again has magnitude 1 as found before for a
combination of traveling and evanescent waves. \ The lowest energy states are
the fully negative energy states and have $n_{-k}(z)+f_{k}$ $n_{k}(z)$ on the
left and $g_{k}n_{-k^{\prime}}(z)$ on the right.

As the potential step size $V_{0}$ increases and enters into the range
$V_{0}\geq2m,$ a new pairing of functions that Klein did not consider begins
to appear. \ These states are discussed Table 2. \ Examining the top of Table
2, the highest energy range $E_{k}>V_{0}+m$ is unchanged from the previous
discussion. \ The second energy range is now expanded to $V_{0}-m\leq
E_{\kappa}\leq V_{0}+m$ and remains, as before, with complete reflection.
\ The new range $m\leq E_{k}\leq V_{0}-m$ pairs wave functions $p_{k}%
(z)+f_{k}$ $p_{-k}(z)$ on the left with transmitted functions $g_{k}%
n_{-k^{\prime}}(z)$ on the right$.$ \ The remaining lower energy ranges and
functions are not changed significantly from the previous discussion. \ This
new set of solutions on both sides in the energy range $m\leq E_{k}\leq
V_{0}-m$ is what Klein did not consider. \ In the following we will solve for
the reflection and transmission amplitudes for those cases with non-zero
probability current density.

For energies above the step, $E_{k}\geq V_{0}+m$ it is straightforward to show
that%
\begin{equation}
\sqrt{m^{2}+k^{2}}=V_{0}+\sqrt{m^{2}+k^{\prime2}}%
\end{equation}

\begin{equation}
f_{k}=\frac{(u_{k}-u_{k}^{\prime})}{(u_{k}+u_{k}^{\prime})}\;,\;g_{k}%
=\frac{2u_{k}}{(u_{k}+u_{k}^{\prime})},
\end{equation}
where
\begin{equation}
u_{k}=\frac{k}{m+\sqrt{m^{2}+k^{2}}}%
\end{equation}
and%
\begin{equation}
u_{k}^{\prime}=\sqrt{\frac{\sqrt{m^{2}+k^{2}}-V_{0}-m}{\sqrt{m^{2}+k^{2}%
}-V_{0}+m}}.
\end{equation}
Using the usual definitions of current density, it is easy to determine that
the reflection coefficient $R$ and the transmission coefficient $T$ are given
by%
\begin{equation}
R=|f_{k}|^{2}=\frac{(u_{k}-u_{k}^{\prime})^{2}}{(u_{k}+u_{k}^{\prime})^{2}}%
\end{equation}
and%
\begin{equation}
T=\frac{u_{k}^{\prime}}{u_{k}}g_{k}^{2}=\frac{4u_{k}u_{k}^{\prime}}%
{(u_{k}+u_{k}^{\prime})^{2}}%
\end{equation}
from which it is easy to see that $R+T=1.$

For the ranges of energies in which the waves on the right (or the left) are
evanescent, we easily find that $T=0$ and $R=1.$

For the case which Klein did not consider, when $V_{0}\geq2m$ similar
calculations yield%
\begin{equation}
\sqrt{m^{2}+k^{2}}=V_{0}-\sqrt{m^{2}+k^{\prime2}}%
\end{equation}%
\begin{equation}
f_{k}=\frac{(u_{k}u_{k}^{\prime}-1)}{(u_{k}u_{k}^{\prime}+1)}\;,\;g_{k}%
=\frac{2u_{k}}{(u_{k}u_{k}^{\prime}+1)}%
\end{equation}

\begin{equation}
u_{k}^{\prime}=\sqrt{\frac{V_{0}-\sqrt{m^{2}+k^{2}}-m}{V_{0}-\sqrt{m^{2}%
+k^{2}}+m}}.
\end{equation}
and $u_{k}$ is defined as before. \ The reflection and transmission
coefficients are:%
\begin{equation}
R=|f_{k}|^{2}=\frac{(u_{k}u_{k}^{\prime}-1)^{2}}{(u_{k}u_{k}^{\prime}+1)^{2}}%
\end{equation}
and%
\begin{equation}
T=\frac{u_{k}^{\prime}}{u_{k}}g_{k}^{2}=\frac{4u_{k}u_{k}^{\prime}}%
{(u_{k}u_{k}^{\prime}+1)^{2}}%
\end{equation}
from which it is also easy to see that $R+T=1.$

From these calculations it is clear that there is no paradox. \ Klein simply
did not match the appropriate solutions in the two regions. \ This mis-match
is what is responsible for the usual assumption of a failure of particle
conservation (usually taken to imply the production of particle hole pairs
near the potential step) that has been so often interpreted as the meaning of
the Klein paradox. \ These analytic solutions show that particle number is
always conserved independent of the size of the step potential. \ The
surprising phenomena is that the barrier is close to being transparent in a
certain energy range if the step height $V_{0}$ is large enough. \ This
behavior of the reflection coefficient is shown in Fig. 1 where the reflection
coefficient $R$ has been simultaneously plotted for three different values of
$V_{0}=0.5m,$ $3m,$ $8m.$ \ In all of these cases there is a region of
complete reflection with an energy width of $2m$ and centered on the value of
$V_{0}.$ \ At higher energies the reflection coefficient decreases with
increasing energy.

The surprising result is that the reflection coefficient decreases in the
energy interval $m\leq E_{k}\leq V_{0}-m.$ and the barrier becomes partially
transparent in this energy range. \ This result is an exact consequence of the
matching of exact solutions in the two regions joined by the boundary
condition. \ On physical grounds this result would seem to be unexpected, but
the potential step has pulled what used to be negative energy states into a
positive energy range and their ability to carry current leads to the partial
transparency of the potential step in this energy range. \ One possible origin
for this surprising phenomena could be with Dirac's original choice for adding
potential energies onto the free particle Dirac Hamiltonian. \ 

\section{Overlap between states of non-interacting and interacting
Hamiltonians}

Another aspect of the behavior of this system can be probed by studying the
overlap of these interacting states in the presence of the potential step with
the free states of the free particle Hamiltonian. \ This is equivalent to
asking how the wave functions are matched in the sudden approximation if the
step potential were instantaneously turned on. \ If there were a production of
extra electron positron pairs by the potential step, it could be expected that
this overlap would be an indicator of such when the potential $V_{0}$ becomes
larger than $2m.$ \ In order to carry out this estimate we would need to
calculate%
\begin{equation}
N=\sum_{k^{\prime\prime},P_{k}}|\langle n_{k^{\prime\prime}}|P_{k}\rangle
|^{2},
\end{equation}
where $n_{k^{\prime\prime}}(z)$ is a negative energy wave function of the zero
potential Hamiltonian spanning both intervals $(-L,0)$ and $(0,L),$ and
$P_{k}(z)$ represents any positive energy state of the Hamiltonian with the
potential step $V_{0}.$ \ And the expression is summed over all positive
energy states with non-zero matrix elements.

It is first important to notice that all of the positive energy states of on
both sides of the potential step, which are above the mass gaps on each side,
will have no contribution to this matrix element because all of these states
are made up of linear combinations of $p_{k}(z)$ and $p_{-k}(z)$ which are
orthogonal to $n_{k^{\prime\prime}}(z)$ in both of the intervals $(-L,0)$ and
$(0,L).$ \ So, the only states that can overlap with the negative energy
states can be:

(1) evanescent states in the mass gaps (which will be shown to be negligable
in the thermodynamic limit, see Appendix I),

(2) the negative energy states $n_{k}(z)+e^{i\phi}n_{-k}(z)$ in the range
$-m\leq E_{k^{\prime}}\leq V_{0}-m$ which are matched with evanescent states.
\ The wave functions for these are of the form%
\begin{equation}
\Psi_{2}(z)=A_{2}((O_{-\kappa}(z)\Theta(-z)+(n_{k^{\prime}}(z)+e^{i\phi
}n_{-k^{\prime}}(z))\Theta(z)),
\end{equation}
and

(3) positive energy states appearing when $V_{0}\geq2m$ which carry current on
both sides%
\begin{equation}
\Psi_{3}(z)=A_{3}((p_{k}(z)+f_{k}p_{-k}(z))\Theta(-z)+(g_{k}n_{-k^{\prime}%
}(z))\Theta(z)).
\end{equation}

Before evaluating these three cases, it is possible to arrive at an intuitive
estimate of this quantity by simply asking from a density of states
perspective how many of these originally negative energy states have been
pulled up from energy $-m$ to positive energies $V_{0}-\sqrt{m^{2}+k^{2}}$ by
the magnitude of $V_{0}.$ \ If we ignore the boundary conditions, the number
of such states would be given by the integral%
\begin{equation}
\frac{L}{2\pi}\int_{0}^{\sqrt{V_{0}(V_{0}+2m)}}dk=\frac{L\sqrt{V_{0}%
(V_{0}+2m)}}{2\pi}.
\end{equation}

Let us first examine case (2). \ The first step is to evaluate the
normalization $A_{2}$ which yields%
\begin{equation}
A_{2}(k)=\frac{1}{\sqrt{L(2(1+u_{k^{\prime}}^{2})+\frac{(1-e^{-\kappa L)}%
}{\kappa L}).}}.
\end{equation}
Neglecting terms of order $1/L$ and smaller and using the orthogonality of
$n_{k^{\prime}}(z)$ and $n_{k^{\prime\prime}}(z)$ it is straightforward to
show that%
\begin{equation}
\langle n_{k^{\prime\prime}}|\Psi_{2}(k)\rangle=\frac{1}{2}(\Delta
_{k^{\prime\prime},k^{\prime}}+e^{i\phi}\Delta_{k^{\prime\prime},-k^{\prime}%
}).
\end{equation}
Now, for each $L$ in the sequence to the thermodynamic limit, it is true that%
\begin{equation}
\Delta_{k^{\prime\prime},k^{\prime}}^{2}=\Delta_{k^{\prime\prime},k^{\prime}}%
\end{equation}
so we obtain%
\begin{equation}
|\langle n_{k^{\prime\prime}}|\Psi_{2}(k)\rangle|^{2}=\frac{1}{4}%
(\Delta_{k^{\prime\prime},k^{\prime}}+\Delta_{k^{\prime\prime},-k^{\prime}}).
\end{equation}
This implies that%
\begin{equation}
N_{2}=\sum_{k^{\prime\prime},k}|\langle n_{k^{\prime\prime}}|\Psi
_{2}(k)\rangle|^{2}=\frac{1}{2}\sum_{k^{\prime}}1=\frac{L}{4\pi}\sqrt
{V_{0}(V_{0}+2m)}%
\end{equation}
for the energy range $-m\leq E_{k^{\prime}}\leq V_{0}-m.$ \ If $V_{0}>2m,$ the
lower limit of the $k^{\prime}$ integration becomes $\sqrt{V_{0}(V_{0}-2m)}$
so that the expression for $N_{2}$ in that case becomes%
\begin{equation}
N_{2}=\frac{L}{4\pi}(\sqrt{V_{0}(V_{0}+2m)}-\sqrt{V_{0}(V_{0}-2m)}).
\end{equation}

Now let us examine case (3). \ In this case the matrix element becomes%
\begin{equation}
\langle n_{k^{\prime\prime}}|\Psi_{3}(k)\rangle=A_{3}(k)g_{k}\int_{0}%
^{L}n_{k^{\prime\prime}}(z)^{\dagger}n_{-k^{\prime}}(z)dz
\end{equation}
where%
\begin{equation}
A_{3}(k)=\frac{1}{\sqrt{(1+u_{k}^{2})(1+f_{k}^{2})+g_{k}^{2}(1+u_{k^{\prime}%
}^{2})}}.
\end{equation}
Substituting for $f_{k}^{2}$ and $g_{k}$ from the eqn. (20) and (21) above, we
find%
\begin{equation}
\langle n_{k^{\prime\prime}}|\Psi_{3}(k)\rangle=\frac{\sqrt{2}u_{k}%
\sqrt{(1+u_{k^{\prime}}^{2})}\Delta_{k^{\prime\prime},-k^{\prime}}}%
{\sqrt{u_{k}^{2}u_{k^{\prime}}^{2}(3+u_{k}^{2})+(1+3u_{k}^{2})}}.
\end{equation}
So, the contribution from this case is%
\begin{equation}
N_{3}=\frac{L}{2\pi}\int_{0}^{\sqrt{V_{0}(V_{0}-2m)}}\frac{2u_{k}%
^{2}(1+u_{k^{\prime}}^{2})}{u_{k}^{2}u_{k^{\prime}}^{2}(3+u_{k}^{2}%
)+(1+3u_{k}^{2})}dk.
\end{equation}

The remaining case 1 is analyzed in Appendix I and gives a result which is not
extensive with the length $L$ and so makes no contribution in the
thermodynamic limit. \ The details and certain aspects of orthogonality in the
thermodynamic limit are presented in Appendix I.

The dependence of $N$ on the barrier height $V_{0}$ is shown in Fig. 2. \ For
small $V_{0}$ the intuitive estimate and the exact result agree. \ When the
potential height becomes greater then $2m,$ the lower limit of Eqn. (33) and
the integral for $N_{3}$ makes their contribution. \ The smooth curve is the
intuitive estimate given by Eqn. (32) and the lower curve with the kink at
$V_{0}=2m$ is the exact result for this quantity. \ Note that the behavior of
this quantity is smooth after the threshold $V_{0}=2m$ and less than the
intuitive estimate.

\section{Transmission through square barriers}

Walker and Gathright\cite{WalkerandGathright} worked out all possible one
dimensional transfer matrices for the non-relativistic Schrodinger equation
across potential discontinuities. \ They constructed transfer matrices for any
arrangement of potential discontinuities by building the transfer matrices out
of products of two different matrices: \ the discontinuity matrix $d$ and the
propagation matrix $P.$ \ All such possible matrices have been worked out for
the Dirac equation and will be presented elsewhere. \ For the one dimensional
Dirac equation\ the discontinuity matrix $d$ is of the form%
\begin{equation}
d(a,b)=\left[
\begin{array}
[c]{cc}%
a+b & a-b\\
a-b & a+b
\end{array}
\right]  ,
\end{equation}
and the parameters $a$ and $b$ are replaced by $1$ or by $u_{k}$ or
$iw_{\kappa}$ or by ratios of these depending on the energy range in the
segment between two discontinuities of the potential. \ The propagation matrix
is identical to those found by Walker and Gathright
\begin{equation}
P(\alpha)=\left[
\begin{array}
[c]{cc}%
e^{\alpha} & 0\\
0 & e^{-\alpha}%
\end{array}
\right]  .
\end{equation}
Once relativistic matrices $d$ and $P$ have been constructed for all possible
pairs of states which are possible at a discontinuity, it is immediate to
construct the transfer matrices that will connect the two solutions at the
ends of a region $l$ where the potential has a constant value $V_{l}.$ \ A
simple, symmetric example of the use of these transfer matrices is a square
barrier whose height is $V_{0}=5.5$ and whose width is $a/L=5/150=1/30.$ \ A
plot of the reflection coefficient $R$ versus the momentum of the initial wave
$k$ is shown in Fig. 3 . \ Notice the Ramsauer minima (peaks in $T$ ) in the
Reflection coefficient.\cite{Ramsauer} \ Notice also that the Reflection
coefficient for this symmetric case behaves much like the step potential if
the potential energy $V_{0}$ is larger than the mass gap width and becomes
semi-transparent at lower energies. \ 

\section{Properties of zero mass solutions of the Dirac equation}

The well known analogy between the band structure of graphene and the energy
spectrum of the massless Dirac equation has received much attention recently.
\ For the non-relativistic band structure calculations the velocity of a band
state represented by the energy $\epsilon_{k}$ is linear in $|k|$ and is given
by $\epsilon_{k}=\pm v|k|$ (in one dimension). \ For electronic band
structures the current density is proportional to the group velocity which is
given by the derivative of the band energies,%
\begin{equation}
v_{k}=\frac{d\epsilon_{k}}{dk},
\end{equation}
and the :"graphene" band structure spectrum will contain both positive and
negative velocities for both positive and negative energies (relative to the
center of the band). \ 

Examining the zero mass eigenstates of the Dirac equation shows that there are
some quite different behaviors than those observed in the non-relativistic
band structure for Graphene. \ The (1+1) Dirac equation has the following
simple form in a constant potential $V_{0}$%
\begin{equation}
\left[
\begin{array}
[c]{cc}%
0 & \frac{d}{idz}\\
\frac{d}{idz} & 0
\end{array}
\right]  \left(
\begin{array}
[c]{c}%
\alpha\\
\beta
\end{array}
\right)  e^{ikz}=(E-V_{0})\left(
\begin{array}
[c]{c}%
\alpha\\
\beta
\end{array}
\right)  e^{ikz}%
\end{equation}
has eigenstates%
\begin{equation}
\Psi_{\pm}(k)=\frac{1}{\sqrt{2}}\left(
\begin{array}
[c]{c}%
1\\
\pm1
\end{array}
\right)  e^{ikz}.
\end{equation}
with correspond to the energies%
\begin{equation}
E-V_{0}=\pm k.
\end{equation}

The probability current densities carried by these states are:%
\begin{equation}
J_{+}=\Psi_{+}^{\dagger}\sigma_{x}\Psi_{+}=\frac{1}{2}\left(
\begin{array}
[c]{cc}%
1 & 1
\end{array}
\right)  \left(
\begin{array}
[c]{cc}%
0 & 1\\
1 & 0
\end{array}
\right)  \left(
\begin{array}
[c]{c}%
1\\
1
\end{array}
\right)  =1
\end{equation}
and%
\begin{equation}
J_{-}=\Psi_{-}^{\dagger}\sigma_{x}\Psi_{-}=-1.
\end{equation}
So the positive direction currents are carried only by positive energies
$E-V_{0}\geq0$ and negative direction currents are carried by $E-V_{0}\leq0.$
\ The massless Dirac equation does not allow negative currents to be carried
by positive energy eigenstates.

If we consider the reflection at a potential step at $z=0$ and apply the
boundary condition we obtain,
\begin{equation}
k=V_{0}+k^{\prime}%
\end{equation}
and%
\begin{equation}
\left(
\begin{array}
[c]{c}%
1\\
1
\end{array}
\right)  +f_{k}\left(
\begin{array}
[c]{c}%
1\\
-1
\end{array}
\right)  =g_{k^{\prime}}\left(
\begin{array}
[c]{c}%
1\\
1
\end{array}
\right)  .
\end{equation}
which yields the equations%
\begin{equation}
1+f_{k}=g_{k^{\prime}}%
\end{equation}%
\begin{equation}
1-f_{k}=g_{k^{\prime}}%
\end{equation}
which implies that $f_{k}=0$ and $g_{k^{\prime}}=1.$ \ This condition implies
that the potential step makes no reflections for a zero mass particle. \ This
condition and the fact that current direction is so strongly associated with
the sign of the energy (relative to $V_{0}$ ) indicates that the analogy
between the graphene bandstructure and the massless Dirac equation is not
completely accurate.

\section{Conclusions}

The Klein Paradox is not a paradox. \ It is simply a mis-application of the
processes by which the solution of piece-wise constant potential differential
equations are constructed. \ When the appropriate wave functions at the same
energies are connected, the reflection and transmission coefficients are
continuous functions of the incident wave vector and always obey the
conservation of particle number. \ The surprising characteristic of these
solutions is the near transparency of the step potential at low energies if
$V_{0}>2m.$ \ This property of the solutions arises from the fact that for
large $V_{0}$ states which were originally at negative energies are now pulled
up into positive energies and it becomes possible for a current to be carried
through the step. \ This result is clearly a property of the solutions of this
Dirac equation. \ The more difficult question is whether this behavior is
physically to be expected. \ This property reflects a choice made by Dirac
when he decided to add a potential energy to his free particle equation. \ He
chose to add it to the $\boldsymbol{\alpha}\cdot\boldsymbol{p}$ as opposed to
adding the potential energy to the mass $m$. \ These questions have been
examined to some extent in other contexts\cite{greiner}. \ 

The connection between the overlap of the negative energy states of the
non-interacting Hamiltonian and the Hamiltonian with the potential step, if
the latter is turned on instantaneously was examined by studying the summation
of the overlap matrix element between positive energy states and the initially
negative energy states. \ It was shown that the accurately determined overlap
was consistent with an intuitive picture that the overlap represented the
number of negative energy states that have been pulled above the energy
$E=-m.$ \ As should be expected because the Klein paradox does not exist,
there is no anomalous behavior of this overlap as the potential step is
increased above the threshold $V_{0}=2m$. \ 

Generalizing the transfer matrices from the non-relativisitic Schrodinger
equation to the Dirac equation allows the treatment of a variety of potential
barriers and steps and, numerically, any smooth potential that can be
approximated by piece-wise constant potentials in short intervals. \ By way of
an example, the case of a square potential barrier was briefly discussed.
\ The presence of Ramsauer resonances and the transparency of the barrier were
found in direct analogy to the results of the step potential.

Finally, zero mass eigenstates of the Dirac equation were examined. \ Positive
current densities were only carried by positive energy states (relative to
$V_{0})$ and negative current densities were only carried by negative energy
states. \ It appears that a step potential at the origin has no effect on
these states. \ Both of these conditions are quite different from the band
structure of Graphene which has stimulated the analogy between that material
and the solutions to the Dirac equation. \ The failure of a step potential to
influence the zero energy states seems to be quite unphysical, and seems again
to be related to Dirac's original choice by which he added the potential
energy to the free particle Dirac equation. \ 

\section{Acknowledgements}

The author acknowledges incisive and helpful conversations with Reiner Grobe
and Charles Su with support from their National Science Foundation Grant, and
helpful conversations with Jay Mancini of Kingsborough College and John Gray
of the Dahlgren Naval Surface Weapons Center.

\section{Appendix I: Overlap between Evanescent and Free Negative Energy
States}

In this appendix the overlap of the negative energy free particle eigenstates
with the evanescent states in the gaps are examined and found to be of order
$1/L.$

As an example, consider a state where the overall energy is in the range
$V_{0}\leq E\leq V_{0}+m.$ \ One of the wave functions in this range is%

\begin{equation}
\Psi_{1}(z)=A_{1}(k)((p_{k}(z)+e^{i\phi}p_{-k}(z))\Theta(-z)+(g_{\kappa
}e^{-\kappa z}\binom{1}{iw_{\kappa}})\Theta(z)), \tag{A-1}%
\end{equation}
where%

\begin{equation}
A_{1}(k)=\frac{1}{\sqrt{L}\sqrt{2(1+u_{k}^{2})+g_{k}^{2}(1+w_{\kappa}%
^{2})\frac{(1-e^{-2\kappa L})}{\kappa L}}}. \tag{A-2}%
\end{equation}
The matrix element, ignoring terms of order $e^{-\kappa L},$ is
\begin{equation}
\langle n_{k^{\prime\prime}}|\Psi_{1}(k)\rangle=\frac{1}{L}\frac
{(u_{k}+iw_{\kappa})}{\sqrt{2}\sqrt{(1+u_{k^{\prime\prime}}^{2})}%
\sqrt{2(1+u_{k}^{2})+\frac{g_{k}^{2}}{\kappa L}(1+w_{\kappa}^{2})}%
(\kappa-ik^{\prime\prime})}. \tag{A-3}%
\end{equation}

The contribution of this matrix element to $N_{1}$ as $L\rightarrow\infty$ is
given by%
\begin{equation}
N_{1}=\sum_{k^{\prime\prime},k}|\langle n_{k^{\prime\prime}}|\Psi_{1}%
\rangle|^{2}=\frac{1}{16\pi^{2}}\int dk^{\prime\prime}\int dk\frac{(u_{k}%
^{2}+w_{\kappa}^{2})}{(1+u_{k^{\prime\prime}}^{2})(1+u_{k}^{2})(\kappa
^{2}+k^{\prime\prime2})}. \tag{A-4}%
\end{equation}
Since this integral is not extensive in $L$ when compared to the other
contributions which are proportional to $L,$ this contribution is negligible
in the thermodynamic limit.

The fact that the overlap of $n_{k^{\prime\prime}}(z)$ with the evanescent
states $O_{\kappa}(z)$ was negligable in the thermodynamic limit raises a
question involving evanescent states in the mass gap and more generally parts
of wave functions within the confines of a square potential barrier $0<z<a<L.$
\ This detail has already been discussed by Arfken and Sneddon in their
discussion of the evolution of the fourier integral from the Fourier series in
the thermodynamic limit. \ In their discussion it is observed that in the
transition $L\rightarrow\infty$ the constant term $a_{0}/2$ of the Fourier
series becomes negligible in the thermodynamic limit.

Similarly, the corresponding relationship between "box normalized" wave
functions in the thermodynamic limit becomes apparent in the theorem for
Hermitian operators that eigenvectors with different eigenvalues must be
orthogonal. \ Consider two states of the type from case 1 with different wave
vectors, and thus different energies%
\begin{equation}
\Psi_{k_{1}}(z)=A_{1}(k_{1})((p_{k_{1}}(z)+e^{i\phi}p_{-k_{1}}(z))\Theta
(-z)+(g_{\kappa_{1}}e^{-\kappa_{1}z}\binom{1}{iw_{\kappa_{1}}})\Theta(z))
\tag{A-5}%
\end{equation}%
\begin{equation}
\Psi_{k_{2}}(z)=A_{2}(k_{2})((p_{k_{2}}(z)+e^{i\phi}p_{-k_{2}}(z))\Theta
(-z)+(g_{\kappa_{2}}e^{-\kappa_{2}z}\binom{1}{iw_{\kappa_{2}}})\Theta(z))
\tag{A-6}%
\end{equation}
and the energies are:%
\begin{equation}
E_{k_{1}}=\sqrt{m^{2}+k_{1}^{2}}\neq\sqrt{m^{2}+k_{2}^{2}}=E_{k_{2}}.
\tag{A-7}%
\end{equation}

If we evaluate the overlap between these two vectors, the integrals on the
left hand interval $(-L,0)$ immediately give zero because of the orthogonality
of the $p_{k_{1}}$ functions. \ This leaves an integral on the interval
$(0,L)$%
\begin{equation}
\langle\Psi_{k_{1}}|\Psi_{k_{2}}\rangle=A_{1}A_{2}(0+\frac{g_{\kappa_{1}%
}g_{\kappa_{2}}(1+w_{\kappa_{1}}w_{\kappa_{2}})}{\sqrt{1+w_{\kappa_{1}}^{2}%
}\sqrt{1+w_{\kappa_{2}}^{2}}(\kappa_{1}+\kappa_{2})}). \tag{A-8}%
\end{equation}
On first reflection this second term is not zero and appears to violate the
orthogonality theorem, but it must be noted that the factors $A_{1}A_{2}$
provide a factor of $1/L$ and this matrix element vanishes in the
thermodynamic limit. \ The orthogonality of the kind of states combined in the
process of joining solutions of piece-wise constant potentials is dominated by
the momentum eigenstates and integrals over smaller intervals can be non-zero
for finite $L,$ but make no contribution in the thermodynamic limit.

\section{Tables}

Table 1: A tabulation of the energy range and the type of wave functions that
are matched at the origin as well as the equation relating the wave vector $k$
on the left and $k^{\prime}$ on the right for small $V_{0}<2m.$ \ In this
example, the two mass gaps have a significant overlap which contains no wave
functions. \

\begin{tabular}
[c]{|c|c|c|c|}\hline
$E$ range & $\psi_{L}(z)$ & $\psi_{R}(z)$ & $k$ and $k^{\prime}$\\\hline
$E>V_{0}+m$ & $p_{k}(z)+f_{k}$ $p_{-k}(z)$ & $g_{k}p_{k^{\prime}}(z)$ &
$\sqrt{m^{2}+k^{2}}=V_{0}+\sqrt{m^{2}+k^{\prime2}}$\\\hline
$m\leq E\leq V_{0}+m$ & $p_{k}(z)+e^{i\phi}p_{-k}(z)$ & $g_{\kappa}O_{\kappa
}(z)$ & $\sqrt{m^{2}+k^{2}}=V_{0}+\sqrt{m^{2}-\kappa^{2}}$\\\hline
$V_{0}-m\leq E\leq m$ & no w.f. & no w.f. & True Mass gap; no states\\\hline
$-m\leq E\leq V_{0}-m$ & $g_{-\kappa}O_{-\kappa}(z)$ & $n_{k^{\prime}%
}(z)+e^{i\phi}n_{-k^{\prime}}(z)$ & $\pm\sqrt{m^{2}-\kappa^{2}}=V_{0}%
-\sqrt{m^{2}+k^{\prime2}}$\\\hline
$E\leq-m$ & $n_{-k}(z)+f_{k}$ $n_{k}(z)$ & $g_{k}n_{-k^{\prime}}(z)$ &
$-\sqrt{m^{2}+k^{2}}=V_{0}-\sqrt{m^{2}+k^{\prime2}}$\\\hline
\end{tabular}

Table 2: A tabulation of the energy range and the type of wave functions that
are matched at the origin as well as the equation relating the wave vector $k$
on the left and $k^{\prime}$ on the right for large $V_{0}>2m.$ The two mass
gaps have a significant separation and the function pairs that Klein ignored
are included in this energy range. \

\begin{tabular}
[c]{|c|c|c|c|}\hline
$E$ range & $\psi_{L}(z)$ & $\psi_{R}(z)$ & $k$ and $k^{\prime}$\\\hline
$E>V_{0}+m$ & $p_{k}(z)+f_{k}$ $p_{-k}(z)$ & $g_{k}p_{k^{\prime}}(z)$ &
$\sqrt{m^{2}+k^{2}}=V_{0}+\sqrt{m^{2}+k^{\prime2}}$\\\hline
$V_{0}\leq E\leq V_{0}+m$ & $p_{k}(z)+e^{i\phi}p_{-k}(z)$ & $g_{\kappa
}O_{\kappa}^{(+)}(z)$ & $\sqrt{m^{2}+k^{2}}=V_{0}+\sqrt{m^{2}-\kappa^{2}}%
$\\\hline
$V_{0}-m\leq E\leq V_{0}$ & $p_{k}(z)+e^{i\phi}p_{-k}(z)$ & $g_{\kappa
}O_{\kappa}^{(-)}(z)$ & $\sqrt{m^{2}+k^{2}}=V_{0}-\sqrt{m^{2}-\kappa^{2}}%
$\\\hline
$m\leq E\leq V_{0}-m$ & $p_{k}(z)+f_{k}$ $p_{-k}(z)$ & $g_{k}n_{-k^{\prime}%
}(z)$ & $\sqrt{m^{2}+k^{2}}=V_{0}-\sqrt{m^{2}+k^{\prime2}}$\\\hline
$0\leq E\leq m$ & $g_{-\kappa}O_{-\kappa}^{(+)}(z)$ & $n_{k^{\prime}%
}(z)+e^{i\phi}n_{-k^{\prime}}(z)$ & $+\sqrt{m^{2}-\kappa^{2}}=V_{0}%
-\sqrt{m^{2}+k^{\prime2}}$\\\hline
$-m\leq E\leq0$ & $g_{-\kappa}O_{-\kappa}^{(-)}(z)$ & $n_{k^{\prime}%
}(z)+e^{i\phi}n_{-k^{\prime}}(z)$ & $-\sqrt{m^{2}-\kappa^{2}}=V_{0}%
-\sqrt{m^{2}+k^{\prime2}}$\\\hline
$E\leq-m$ & $n_{-k}(z)+f_{k}$ $n_{k}(z)$ & $g_{k}n_{-k^{\prime}}(z)$ &
$-\sqrt{m^{2}+k^{2}}=V_{0}-\sqrt{m^{2}+k^{\prime2}}$\\\hline
\end{tabular}

\section{Figure Captions}

Fig. 1: Three independent plots of the reflection coefficient R versus the
wave vector $k$ of the incident wave for three different values of
$V_{0}=0.5,$ 3.0,8.0 rest masses are plotted together. \ Each curve can be
identified by the value of $k$ at the center of the $R=1$ plateau.

\bigskip

Fig. 2: Two curves, one approximate and one exact, for the overlap $N$ between
the positive energy states in the presence of the step potential and free
negative energy states in the absence of the step potential. \ The smooth
curve is the "intuitive" $N$ derived from the density of negative energy
states shifted up by the potential step. \ The lower curve with the break at
$V_{0}=2$ is the complete calculation of $N$ as a function of $V_{0}$ for
states satisfying the boundary condition. \ Note that the exact curve does not
indicate excessive overlap (electron-positron) pairs above the threshold
$V_{0}=2m.$

Fig. 3: A plot of the reflection coefficient $R$ versus the inicident wave
momentum $k$ for a square barrier with height $V_{0}=5.5$ and width
$a/L=1/30.$ \ Notice the prominent Ramsauer minima.

\pagebreak

Figure 1%

\begin{figure}
[th]
\begin{center}
\includegraphics[
height=2.7691in,
width=4.2134in
]%
{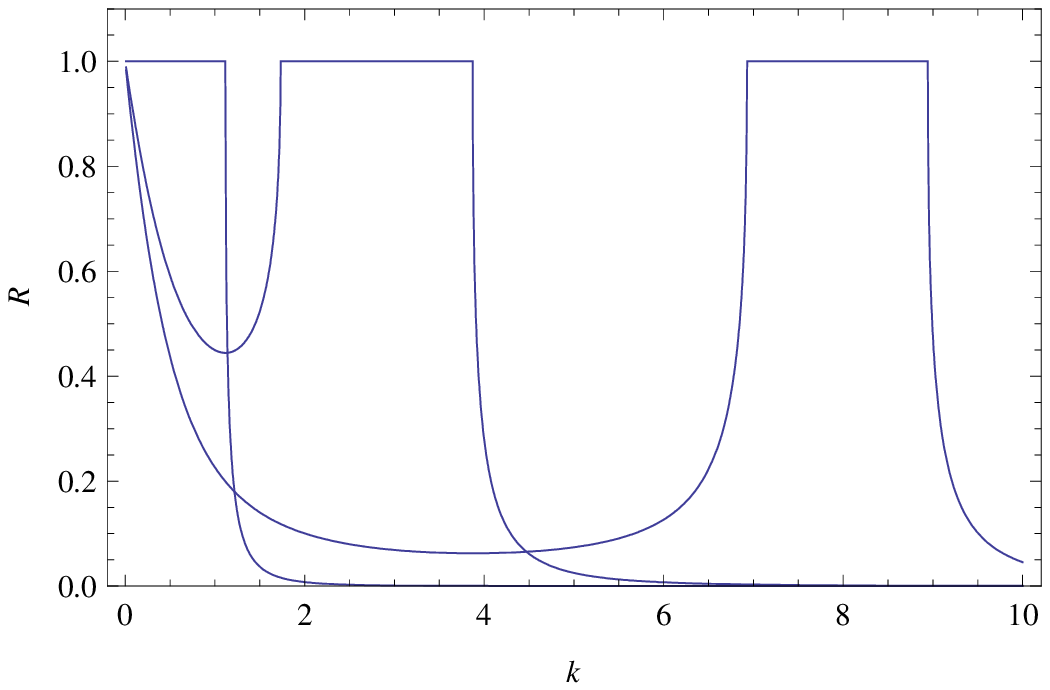}%
\end{center}
\end{figure}

Figure 2%

\begin{figure}
[bh]
\begin{center}
\includegraphics[
height=2.7691in,
width=4.2134in
]%
{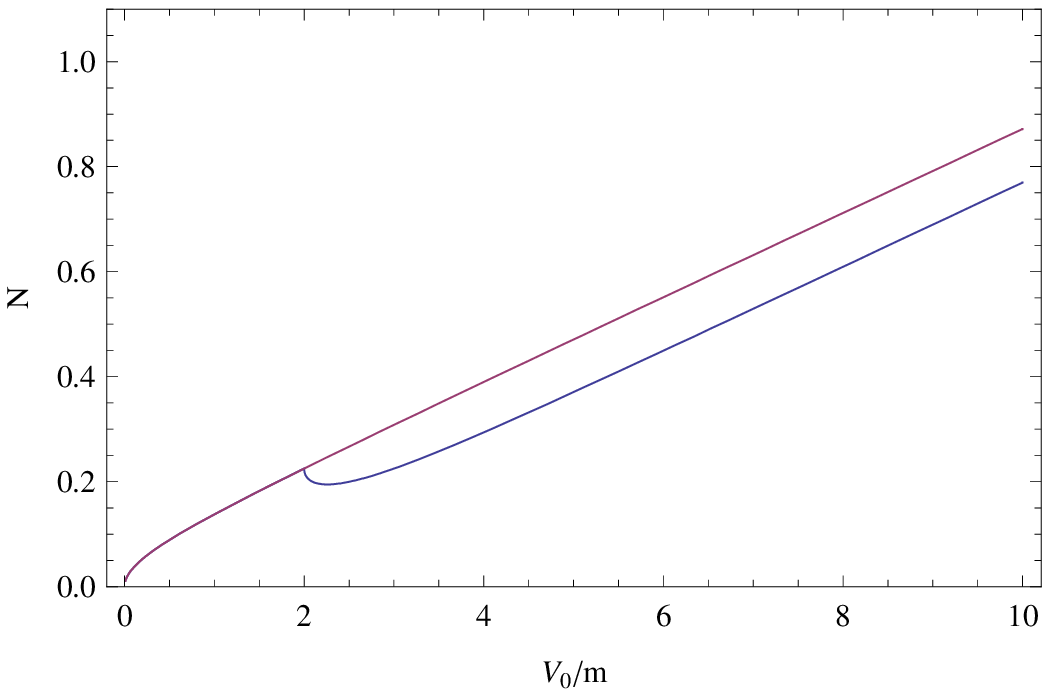}%
\end{center}
\end{figure}

\pagebreak

Figure 3%

\begin{figure}
[th]
\begin{center}
\includegraphics[
height=2.7691in,
width=4.2134in
]%
{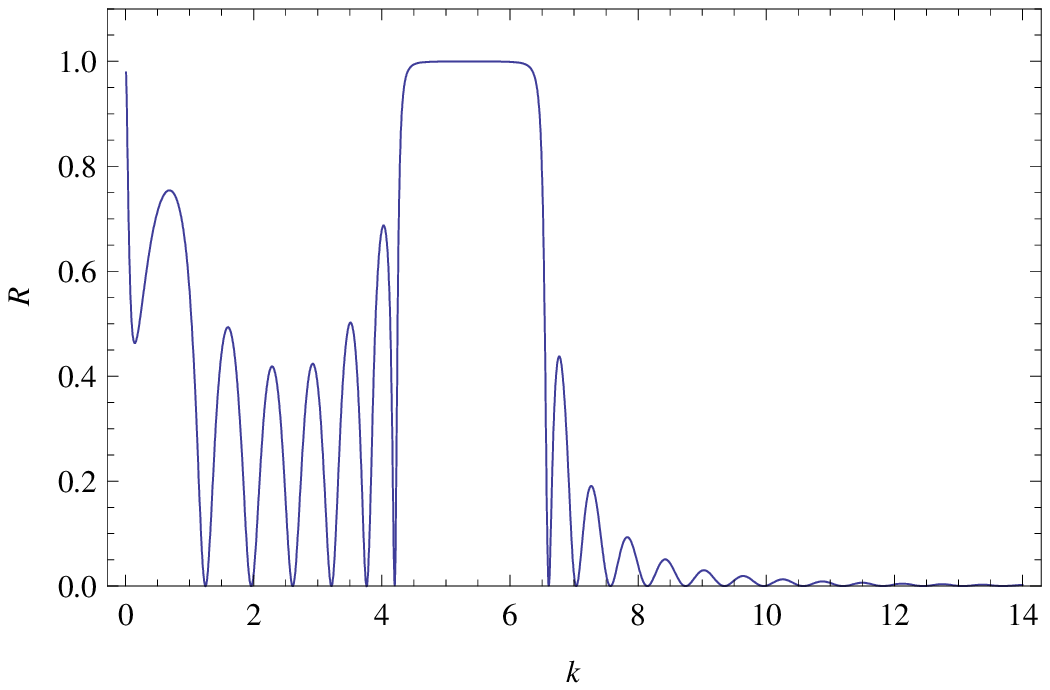}%
\end{center}
\end{figure}

\end{document}